# From Maneuver to Mishap: A Systematic Literature Review on U-Turn Safety Risks

Syed Aaqib Javed*
Ingram School of Engineering
Texas State University
601 University Drive, San Marcos, Texas 78666
Email: aaqib.ce@txstate.edu
ORCID: 0000-0001-5866-035X

Anannya Ghosh Tusti
Ingram School of Engineering
Texas State University
601 University Drive, San Marcos, Texas 78666
Email: gpk30@txstate.edu
ORCID: 0009-0001-5102-5182

Biplov Pandey
Ingram School of Engineering
Texas State University
601 University Drive, San Marcos, Texas 78666
Email: iub14@txstate.edu
ORCID: 0009-0003-2166-9387

Subasish Das, Ph.D.
Ingram School of Engineering
Texas State University
601 University Drive, San Marcos, Texas 78666
Email: subasish@txstate.edu
ORCID: 0000-0002-1671-2753

*Corresponding author

**ABSTRACT**
Understanding the impacts of U-turn configurations on intersection safety and traffic operations is essential for developing effective strategies to enhance road safety and efficiency. Extensive research has been conducted to investigate the role of geometric designs, driver behavior, and advanced technologies in mitigating crash risks and improving traffic flow at U-turn facilities. By synthesizing this collective body of work through the guidelines of Preferred Reporting Items for Systematic Reviews and Meta-Analyses (PRISMA), this paper provides a valuable resource for transportation professionals, policymakers, and researchers seeking evidence-based solutions. This systematic review draws on studies from diverse traffic environments and regional contexts, focusing on innovative design interventions, such as restricted crossing U-turns (RCUTs) and median U-turn intersections (MUTs), as well as integrated strategies leveraging technological advancements. By presenting a comprehensive analysis of U-turn-related challenges and opportunities, this review contributes to advancing transportation safety research and guiding the development of adaptive strategies tailored to varied traffic conditions and evolving technologies.

## INTRODUCTION

U-turn bays near intersections pose significant hazards to urban traffic safety due to the frequent interactions between mixed straight-moving and U-turning traffic flows. These conflict points create an elevated risk of crashes, particularly in high-traffic or unsignalized environments, where drivers often misjudge the speed, distance, or intentions of oncoming vehicles. Studies have reported such risks across countries like Thailand (Meel et al., 2017), India (Mishra et al., 2022), and the United States (FHWA, 2023), highlighting the global nature of this issue. Despite traffic laws giving priority to straight-moving vehicles, conflicts with U-turning vehicles remain common due to factors such as visibility obstructions and difficulties in predicting future trajectories of vehicles in shared U-turn and left-turn lanes. These risks make it imperative to improve the safety of U-turn bays through better design, real-time monitoring, and driver behavior modeling.

According to FHWA, U-turn and left-turn maneuvers at unsignalized median openings have a crash rate of 0.41 crashes per opening per year in urban areas and 0.20 in rural areas (FHWA, 2023). Similarly, studies in India indicate a significant number of crashes and fatalities associated with U-turn movements, further emphasizing the need for interventions to improve safety at these locations (Mishra and Pulugurtha, 2022). These risks are compounded in low- and middle-income countries, where traffic conditions are highly mixed, often lacking clear lane discipline or sufficient signage. The resulting unpredictability of driver intentions creates a dangerous environment for both human-driven and autonomous vehicles, highlighting the need for innovative solutions.

Innovative intersections have emerged as an effective response to the safety and operational challenges posed by conventional designs. Configurations such as median U-turn intersections (MUTs), restricted crossing U-turns (RCUTs), Jughandle, Bowtie, Thru cut, right-turn followed by U-turn (RTUT), and Unconventional Median U-turn (UMU) aim to streamline traffic movements by reducing conflict points and optimizing vehicle trajectories. Studies like NCHRP (2004) have demonstrated how directional median openings and non-traversable medians improve safety by channelizing traffic and minimizing collision risks. Additionally, Schroeder et al. (2024) highlighted how integrating advanced geometric layouts with intelligent systems, including connected vehicle technologies, enhances traffic efficiency and accommodates multimodal users such as pedestrians and bicyclists. These designs also address the specific challenges of unsignalized and urban contexts, offering adaptive solutions that respond dynamically to varying traffic conditions.

Previous efforts to address U-turn safety have focused on analyzing crash risks (Kay, 2022; Kronprasert et al., 2020; Shi et al., 2023), optimizing lane design and signal control (Hu et al., 2022; Jovanovic et al., 2023), and modeling vehicle behavior for safety systems (Choi and Hong, 2021; Gao et al., 2022). Additionally, U-turning behavior has been studied using car-following and gap-acceptance models, which capture longitudinal vehicle behavior (Gupta et al., 2018; Wolelaw et al., 2022). However, these studies often fail to consider the dynamic interactions between multiple vehicles in real-world conditions, where driver intentions can change rapidly due to situational factors. Furthermore, research in this domain has primarily relied on traditional crash data analysis, which is reactive and limited by issues such as underreporting and long observation periods (Paul and Ghosh, 2021). This creates a critical gap in the understanding of U-turn safety, particularly in high-risk areas such as unsignalized intersections and urban environments with heavy traffic.

A comprehensive review of U-turn safety is essential to resolve inconsistencies in existing research and guide future safety interventions. The diversity in methodologies and regional traffic conditions across previous studies has led to varying conclusions, making it difficult to establish universally applicable safety strategies. By systematically synthesizing past research, this review aims to offer a consolidated understanding of crash risks, operational challenges, and driver behavior while identifying key gaps that require further investigation. Additionally, it assesses a range of interventions, including geometric design enhancements and advancements in traffic management technologies, to provide evidence-based recommendations for mitigating U-turn-related crashes. As urbanization accelerates and autonomous vehicle technologies continue to evolve, these insights will play a crucial role in shaping adaptive and integrated safety strategies for both conventional and automated traffic systems.

Many studies have been conducted to analyze U-turn safety, examining issues such as crash frequency and severity, driver decision-making, roadway design and geometric adjustments, application of advanced technologies, and integration of multiple approaches to enhance safety and efficiency. By synthesizing this collective body of work, this review aims to provide researchers, policymakers, and practitioners evidence-based solutions to enhance safety and efficiency at U-turn facilities. To ensure a comprehensive and systematic investigation of the available literature, this study follows the PRISMA guidelines, enabling a methodical search and evaluation of relevant studies. By adhering to these guidelines, the review enhances the rigor and reliability of its findings, contributing to the existing knowledge base and informing future research and policy development in U-turn safety.

## METHODOLOGY

This study adopted the PRISMA guidelines to systematically and comprehensively analyze the available literature on U-turn safety, ensuring a rigorous and dependable research process. The adoption of the PRISMA framework enhanced transparency and reporting quality, involving detailed search and evaluation of relevant studies. The PRISMA guideline, a widely recognized set of recommendations developed in 2009, is designed to standardize the approach for conducting systematic reviews, facilitating replication and critical appraisal of the work. It comprises 27 checklist items and a flowchart guiding researchers through comprehensive literature searches, study selection, data extraction, quality assessment, and result synthesis (Moher et al., 2009). By adhering to these guidelines, the study ensured the rigor and reproducibility of the review, promoting the synthesis of reliable evidence to inform road safety interventions and policymaking.

Additionally, the eligibility criteria for the systematic review were defined using the SPIDER framework, which focuses on qualitative and mixed-methods research designs. The SPIDER framework, developed by Cooke et al. (2012), provides a systematic approach to defining the key components for inclusion and exclusion criteria, ensuring the selection of relevant studies aligned with the research objectives. The framework consists of five components: Sample, Phenomenon of Interest, Design, Evaluation, and Research Type. This method was chosen due to its compatibility with qualitative research and its efficiency in accommodating resource and time constraints (Methley et al., 2014). The following SPIDER criteria were employed for this systematic review:

- **Sample**: The review included studies investigating U-turn safety, focusing on crash occurrence, frequency, and contributing factors. The sample covered studies from various geographic locations, road types, and traffic conditions.
- **Phenomenon of Interest**: The review targeted studies examining factors affecting U-turn safety, including roadway design, driver behavior, and environmental influences. Studies published between 2000 and 2024 were considered to provide a comprehensive analysis of contemporary research.
- **Design**: The review included studies employing quantitative, qualitative, and mixed-methods research designs, encompassing experimental, observational, and case studies related to U-turn safety.
- **Evaluation**: The studies were evaluated based on their examination of U-turn safety, including statistical analyses, modeling approaches, and qualitative findings.
- **Research Type**: Publications from a variety of sources—including peer-reviewed journal articles, conference papers, technical reports, and dissertations—related to U-turn safety were reviewed.

The research team employed several databases, namely TRID, Scopus, Web of Science, and Google Scholar, to carry out a comprehensive literature search. To ensure extensive coverage of relevant literature pertaining to U-turn safety, a combination of search terms was used, such as "U-turn," "traffic safety," "crash," "collision," and "accident," along with their variations. For the systematic review, the PRISMA approach was employed, as depicted in Figure 1. Initially, the study team screened the titles and abstracts of the retrieved literature to assess their alignment with the SPIDER framework's criteria.

The full texts of potentially relevant studies were evaluated to determine their eligibility for inclusion in the systematic review on U-turn safety. The search focused on literature published in English between 2000 and 2024. A total of 1,234 records were identified through database searches, and an additional 43 records were sourced through citation tracking. After removing 260 duplicate records, 974 studies were screened by title and abstract, resulting in 346 full-text studies assessed for eligibility. Among these, 298 studies were excluded for reasons such as lack of focus on U-turn safety, publication before 2000, or being non-English texts. An additional 39 full-text studies identified via other methods were assessed, with 35 studies excluded based on similar criteria. Ultimately, 48 studies from databases and 4 from other sources met the inclusion criteria, culminating in a total of 52 studies included in the systematic review.

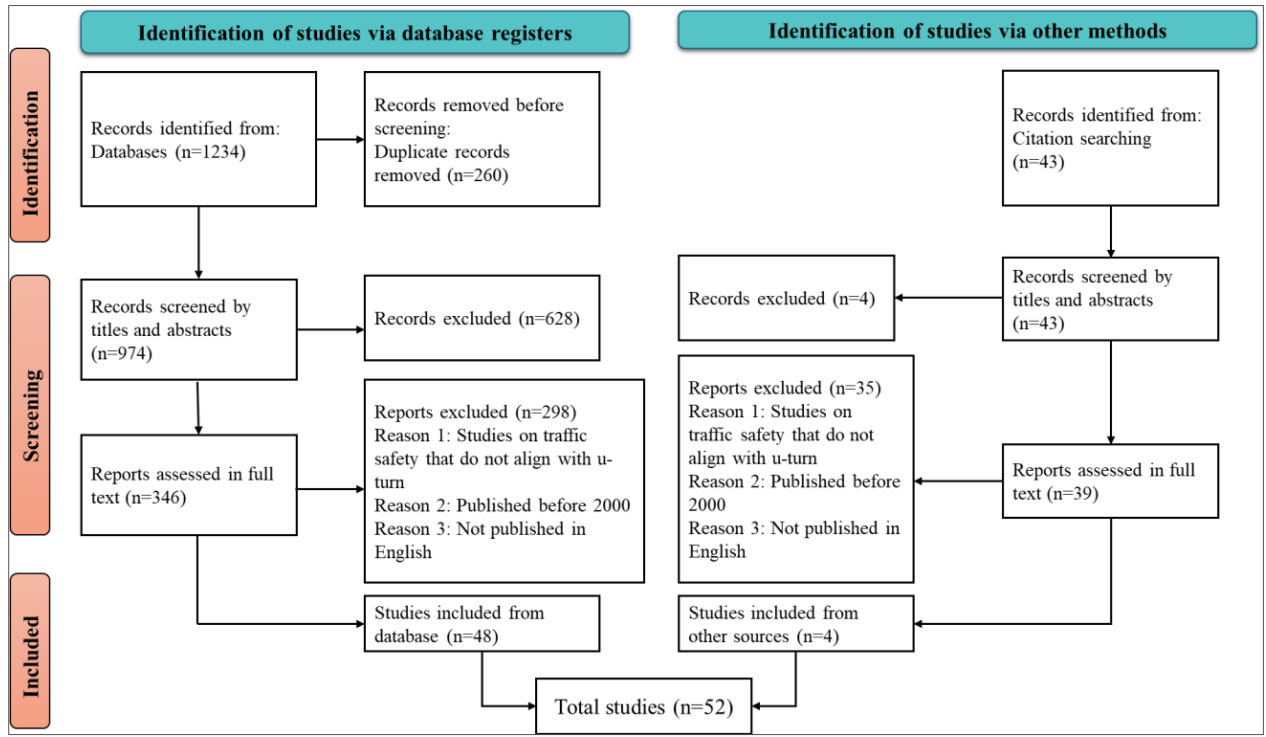

**Figure 1. Flow chart presenting the process of searching and selecting literature.**

This study introduces a structured framework categorizing research into five key focus areas: safety and operations, driver behavior, design interventions, technological innovations, and integrated strategies. Safety and operations explore the impacts of U-turn designs on crash risks and traffic flow, while driver behavior examines compliance and decision-making. Design interventions focus on geometric adjustments to reduce conflict points, and technological innovations highlight advancements like IoT systems and adaptive controls. Integrated strategies synthesize these elements into holistic approaches. Figure 2 illustrates these categories, providing a clear framework for understanding U-turn safety studies and guiding future research and interventions.

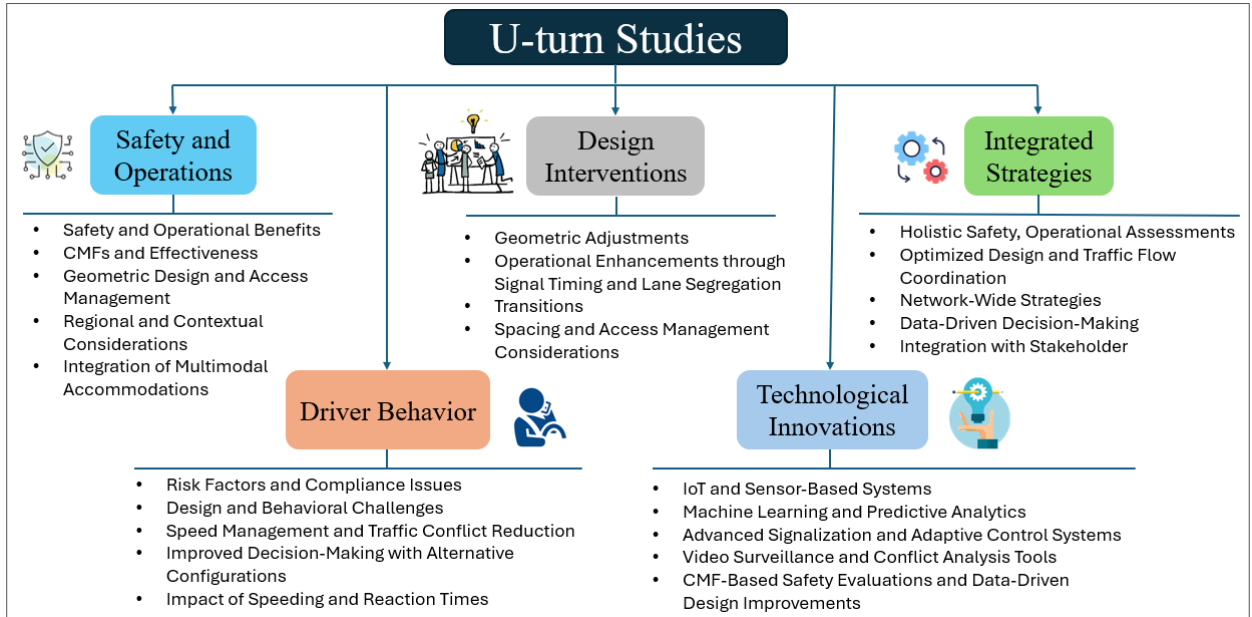

Figure 2. Systematic Classification of Studies into Relevant Categories.

**KEY TOPICS**
**Alternative/Innovative Intersections**
Alternative or innovative intersections are designed to improve traffic flow, enhance safety, and optimize roadway efficiency by addressing the limitations of conventional intersections. Traditional four-leg intersections experience congestion, excessive delay, and high crash rates due to frequent conflict points, particularly at left-turn movements. Alternative designs, such as MUT, RCUT, Jughandles, Thru-Cuts, Bowties, RTUT, and UMU aim to mitigate these challenges by redirecting vehicle movements to reduce direct conflicts and simplify signal phasing. These designs enhance operational performance by minimizing the number of signal phases, reducing cycle lengths, and improving traffic progression along corridors. Additionally, these intersections improve safety by significantly lowering conflict points and eliminating high-risk movements, such as left-turn crashes, which are among the most severe collision types. The implementation of alternative intersections is particularly beneficial in urban and suburban settings where high traffic volumes and multimodal demands necessitate more efficient roadway designs. By incorporating elements like roundabouts, channelized turn lanes, and dedicated pedestrian crossings, these intersections promote a balanced integration of vehicles, pedestrian, and bicycle traffic while accommodating the needs of freight and transit vehicles. The types of alternative intersections are illustrated in Figure 3 and described in the following subsections. The summary of the characteristics of the alternative intersection types are also represented in Table 3.

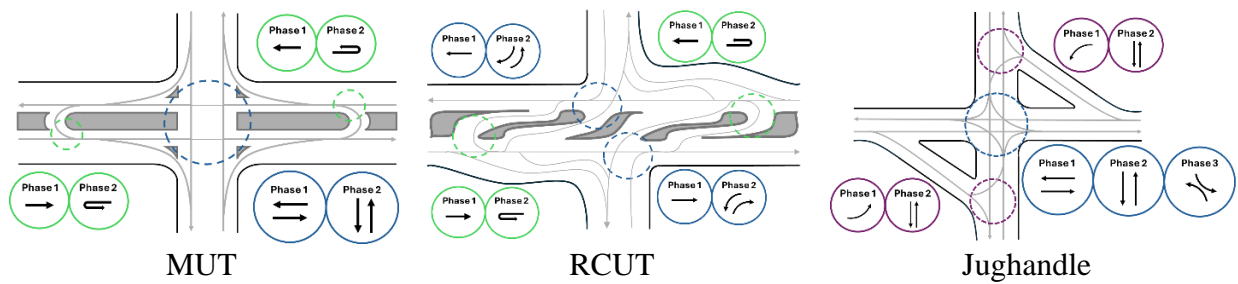

MUT          RCUT          Jughandle

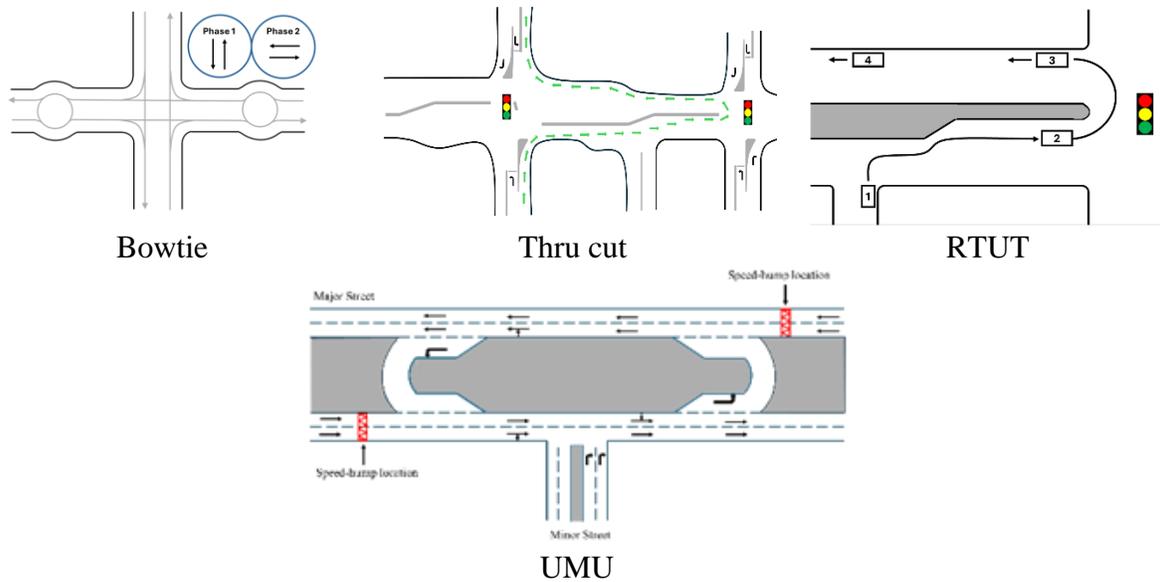

| | | |
|---|---|---|
| Bowtie | Thru cut | RTUT |

UMU

**Figure 3 Types of alternative intersections**

*MUT*

MUT intersections are designed to improve traffic flow and safety by eliminating direct left-turn movements at an intersection. Instead of allowing left turns at the main intersection, MUTs redirect vehicles to proceed straight through the intersection, make a U-turn at a downstream median crossover, and then return to complete the desired movement. This design reduces conflict points, simplifies signal phasing, and can handle high traffic volumes efficiently, particularly in suburban and urban settings with limited right-of-way. The MUT design is effective in reducing severe angle crashes and improving travel time reliability for through and turning movements. Furthermore, the implementation of MUTs aligns well with connected and autonomous vehicle systems due to its predictable traffic flow patterns and minimal critical signal phases (Cunningham et al., 2023; Schroeder et al., 2024a). The CMF calculated in different studies for MUT is highlighted in Table 1.

**Table 1 MUT CMFs**

| Study Name | Crash Type | CMF | Roadway Condition |
|---|---|---|---|
| (Kay et al., 2022) | Fatal and injury crashes | 0.438 | Undivided (two-lane two-way) |
| | Property damage only crashes | 0.035 | Undivided (two-lane two-way) |
| | Fatal and injury crashes | - | Divided (four-lane boulevard) |
| | Property damage only crashes | 1.325 | Divided (four-lane boulevard) |
| (Al-Omari et al., 2020a) | All | 0.633 | Urban and suburban |
| | K, A, B, C | 0.773 | Urban and Suburban |
| | A, B, C | 0.755 | Urban and Suburban |
| | O | 0.598 | Urban and Suburban |
| | Angle | 0.683 | Urban and Suburban |

| | Rear-End | 0.526 | Urban and Suburban |

Note: K = Fatal Injury, A = Incapacitating Injury, B = Non-incapacitating Injury, C = Possible Injury

*RCUT*

RCUT intersections, also known as superstreets, are innovative intersection designs aimed at improving traffic safety and operational efficiency by reducing conflict points and simplifying signal phasing. Unlike traditional intersections where all movements, including left-turns and through movements from minor streets, occur at the primary crossing, RCUTs restrict these movements and redirect them to downstream or upstream U-turns. This design eliminates direct left-turns and through movements from minor approaches, requiring drivers to make a right turn followed by a U-turn. By reconfiguring the flow of traffic, RCUTs minimize the number of critical signal phases, often reducing them to two, which decreases overall delay for major-road traffic and improves progression.

RCUTs are particularly effective in managing high traffic volumes on major arterials while addressing safety concerns. By significantly lowering the potential for severe conflict types such as angle and left-turn crashes, they provide a cost-effective and safer solution for both urban and rural areas. Studies have shown that RCUTs can reduce severe crashes by up to 35% and decrease total crashes by as much as 20% in specific implementations (Cunningham et al., 2023; Schroeder et al., 2024a). Additionally, RCUT designs offer operational benefits, such as improved travel time reliability and better traffic flow. They also accommodate multimodal considerations, including pedestrian crossings, with features like median refuges to enhance safety. These intersections are particularly suitable for locations with limited right-of-way, constrained budgets, or where traditional intersection designs are insufficient for managing traffic demand. The CMF calculated in different studies for RCUT is highlighted in Table 2.

**Table 2 RCUT CMFs**

| Study Name | Crash Type | CMF | Roadway Condition |
|---|---|---|---|
| (Mishra and Pulugurtha, 2021) | All | 0.301 | Rural |
| | K, A, B, C | 0.212 | Rural |
| | All | 0.351 | Suburban |
| | K, A, B, C | 0.266 | Suburban |
| (Sun et al., 2019a) | All | 0.42 | All |
| (Al-Omari et al., 2020a) | All | 0.7632 | Urban and Suburban |
| | K, A, B, C | 0.5669 | Urban and Suburban |
| | Angle | 0.5854 | Urban and Suburban |
| | Head on | 0.0667 | Urban and Suburban |
| | Rear end | 0.7511 | Urban and Suburban |
| | O | 0.8414 | Urban and Suburban |

Note: K = Fatal Injury, A =Incapacitating Injury, B = Non-incapacitating Injury, C = Possible Injury

*Thru-Cut*
At a THRU-CUT intersection, vehicles from the major street can make direct left, right, and through movements, while vehicles from the minor street can execute direct left or right turns. Minor street through movements are redirected to the major street, requiring drivers to make a U-turn or left turn at an adjacent cross street. This design eliminates movement conflicts on the minor street, thereby reducing angle crashes and improving overall safety. Pedestrians can cross the intersection using traditional crosswalks or Z-crossings, which offer a diagonal path across the major street and include median refuge islands for enhanced safety. Bicyclists are accommodated through marked bike lanes, pedestrian crosswalks, or vehicular turning paths, ensuring accessibility for all users.

The THRU-CUT design enhances traffic flow by simplifying signal phasing, typically requiring two or three critical phases depending on the pedestrian facility's layout. This reduction in delays helps mitigate rear-end collisions on the minor street while improving corridor progression for the major street. Although redirected movements increase right-turn volumes, potentially creating pedestrian conflicts, these risks can be mitigated through measures like prohibiting right turns on red or incorporating leading pedestrian intervals. Historically implemented in states like Virginia, Maryland, and North Carolina, the THRU-CUT has proven to be a cost-effective solution for intersections with low minor street through volumes, such as those near residential areas or shopping centers (Cunningham et al., 2023; Hughes et al., 2010; Luo et al., 2024; Schroeder et al., 2024a, 2021; UDOT, 2011).

*Bowtie / Teardrop*
The bowtie intersection is a modern traffic design that eliminates conventional left-turn movements by directing vehicles to nearby roundabouts, where U-turns are executed. This design reduces conflict points, enhances safety, and improves traffic flow. The main intersection supports through and right-turn movements, while left turns are redirected to adjacent roundabouts, reducing crossing conflicts from 16 to 4 and total conflict points to 20, compared to 32 in conventional intersections. The streamlined two-critical-phase signal operation minimizes delays and optimizes pedestrian crossing through shorter cycle lengths. Additionally, shared-use paths ensure safe passage for pedestrians and bicyclists, though these users may encounter increased conflicts with right-turning vehicles. This design is well-suited for intersections with moderate to heavy through traffic and low to moderate left-turn volumes, particularly in urban and suburban settings with limited medians.

The bowtie intersection's operational efficiency stems from its ability to manage traffic flow with minimal disruption. Roundabouts promote slower vehicle speeds, granting drivers more time to react and improving pedestrian and bicyclist safety. Freight vehicles benefit from roundabouts designed with truck aprons, accommodating larger dimensions. However, designers must ensure adequate spacing between the roundabouts and the main intersection to prevent queue spillback. Multimodal facilities include sidewalks, bicycle lanes, and accessible crossings, addressing diverse user needs. Although left-turn demands below 20% of approach traffic make bowtie intersections viable, their implementation requires careful consideration of surrounding land use and right-of-way availability. Furthermore, the absence of median barriers reduces construction complexity while providing opportunities for enhanced corridor progression and connectivity with nearby intersections(Cunningham et al., 2023; Fitzpatrick et al., 2005; Schroeder et al., 2024a).

*Jughandle*

The jughandle U-turn is an innovative intersection design aimed at improving traffic flow and safety by redirecting left-turning vehicles to a designated ramp located either before or after the main intersection. This at-grade design removes the need for left-turn lanes at the primary intersection, thereby reducing congestion and simplifying signal phases. The ramps, positioned in one or more quadrants, facilitate left turns onto minor streets, allowing vehicles to merge back into the main intersection for their desired movement. Jughandles are particularly effective for intersections with high through traffic on the major road and low-to-moderate left-turn volumes on the minor road. These designs often operate with two or three critical signal phases, offering reduced delays compared to conventional four-leg intersections. The inclusion of shared-use paths ensures safe access for pedestrians and cyclists, although these users may experience increased conflict with higher right-turn volumes at ramp junctions. Transit stops, typically located nearside or within the ramp, provide convenient access while minimizing interference with traffic flow.

Jughandle U-turns enhance safety by reducing intersection conflict points from 32 in a conventional design to 24, with significant reductions in angle collisions. The design eliminates left-turn conflicts at the primary intersection while addressing the need for clear signage to mitigate potential driver confusion. However, increased right-turn conflicts with pedestrians and bicyclists at the ramp highlight the need for appropriate multimodal accommodations. Freight vehicles benefit from wide turning lanes and traversable aprons at the ramp junctions, ensuring maneuverability. CMF values for jughandle intersections show a reduction in conflict points and a corresponding decrease in severe crashes, particularly angle collisions. The design's applicability is influenced by corridor spacing, right-of-way availability, and traffic patterns, making it most suitable for suburban or urban intersections with moderate traffic volumes (Cunningham et al., 2023; NJDOT, 2015; Schroeder et al., 2024a).

*RTUT*

RTUT maneuver is a widely adopted alternative to direct left turns (DLT), designed to enhance safety and traffic operations on arterials and access-managed corridors. Instead of executing a direct left turn, vehicles first make a right turn onto the major road, then perform a U-turn at a designated median opening or signalized intersection. This approach significantly reduces conflict points and crash severity, particularly at unsignalized intersections, where gap acceptance for left turns can be challenging. By eliminating direct left-turn movements, RTUT improves traffic flow, minimizes delays, and simplifies signal operations at key intersections. Studies have demonstrated that RTUT reduces the likelihood of severe angle crashes compared to DLT while maintaining comparable or lower overall conflict rates. Although RTUT can introduce additional travel distance, it compensates with smoother operations and improved safety by removing high-risk left-turn crossings.

From a design perspective, adequate median width, spacing between the right turn and U-turn location, and appropriate lane configurations are crucial to ensuring RTUT's effectiveness. Freight vehicles and larger vehicles require wide turn radii and well-designed U-turn bays to accommodate their maneuverability, while pedestrian safety considerations must be addressed at U-turn locations to mitigate conflicts. RTUT has been extensively implemented in states like Florida and Michigan as part of access management strategies, demonstrating its effectiveness in reducing crash risks and optimizing roadway efficiency. However, its success depends on factors such as traffic volume, median availability, and intersection spacing. With the continued emphasis on improving arterial safety and efficiency, RTUT remains a proven and adaptable solution for managing left-turn conflicts in both urban and suburban settings (Lu, 2001; Pirinccioglu et al., 2006).

*UMU*

The UMU is a widely utilized intersection design in various countries, particularly in Egypt, where it is primarily used at unsignalized three-leg and four-leg intersections. The UMU eliminates direct crossing conflicts by replacing traditional left-turn and through movements with right turns followed by U-turns. This design is particularly effective for intersections with low-to-moderate traffic volumes, as it reduces intersection delays and simplifies traffic flow. For traffic volumes under 1250 vehicles/hour per approach, UMUs demonstrate lower delays than signalized intersections, although they perform poorly under high traffic conditions.

In terms of safety performance, UMUs reduce the total number of conflict points compared to conventional intersections. Studies have indicated that they can lower crash risks by 20–50% in comparison to traditional designs. However, the efficiency of UMUs is dependent on several factors, including median width, spacing of U-turn crossovers, and traffic composition. The design is less suitable for intersections with high left-turn demands or limited right-of-way. Additionally, UMUs often require more space for U-turn crossovers, which can be challenging in urban areas. Despite these limitations, UMUs are recognized as an effective strategy for improving traffic safety and operations in specific contexts (Shahdah and Azam, 2021).

**Table 3. Characteristics of Different Types of U-turns**

| U-turn focused Alternative Intersection | Alternative Names | Key Design Feature | Conflict Points |
|---|---|---|---|
| MUT | Michigan Left, Boulevard Left, Indirect Left Turn, ThrU Turn, P-Turn, Median U-turn crossover, Boulevard turnaround, Michigan loon | • Removes direct left turns at intersections, redirecting vehicles to make a U-turn downstream.<br>• Reduces conflict points by eliminating left-turn phases, simplifying traffic signals.<br>• Suitable for corridors with moderate to high traffic volumes and limited right-of-way. | 16 |
| RCUT | Superstreet, J-Turn, Reduced Conflict Intersection (RCI), High-T Intersection, Synchronized street intersection | • Left-turns and through movements from minor streets are replaced with right turns followed by a U-turn.<br>• Improves safety by reducing crossing conflicts, particularly at high-speed roads.<br>• Applicable in areas with significant traffic volume disparities between major and minor streets. | 14 |
| Jughandle | Jersey Left, Forward Jughandle, Near-Side Jughandle, Reverse Jughandle, Far-Side Jughandle | • Redirects left-turning vehicles via ramps located before or after the main intersection.<br>• Reduces congestion by eliminating left-turn lanes and simplifying signal phasing.<br>• Provides shared-use paths for pedestrians and cyclists, though increased right-turn conflicts may arise. | 24 |
| Thru cut | Hamburger intersection, Through-about intersection | • Eliminates through movements on minor streets, redirecting vehicles to adjacent roads for completion.<br>• Reduces signal phases and simplifies intersection operations.<br>• Best suited for intersections with low through traffic on minor streets. | 20 |
| Bowtie | Teardrop Roundabout | • Redirects left turn to adjacent roundabouts for U-turns, minimizing left-turn conflicts.<br>• Operates with a two-critical-phase signal design, reducing pedestrian and vehicle delays.<br>• Ideal for urban settings with moderate through traffic and constrained medians. | 16 |
| RTUT | Florida Left, Directional Median Opening | • Eliminates direct left turns from driveways and minor streets, redirecting vehicles to make a right turn followed by a U-turn at a downstream median.<br>• Reduces conflict points by removing left-turn movements at intersections.<br>• Enhances traffic flow and safety by addressing gap acceptance challenges and simplifying left-turn maneuvers.<br>• Effective for access-managed arterials with high through traffic and moderate left-turn volumes.<br>• Requires adequate spacing between the right-turn and U-turn location for safe lane changes. | 18 |
| UMU | Unconventional Median U-turn (UMUT) | • Eliminate direct crossing conflicts by replacing traditional left-turn movements with right turns followed by U-turns.<br>• Effective for low-to-moderate traffic volumes.<br>• Requires wider medians for U-turn crossovers.<br>• Reduces crash risks by 20–50% compared to conventional designs. | 14 |

**Safety and Operations**
The evaluation of safety and operational impacts of U-turn configurations, including RCUTs and median U-turns, has highlighted significant improvements in reducing conflict points, enhancing traffic flow, and mitigating crashes. Dixon et al. (2018) analyzed Texas U-turn configurations at diamond interchanges, reporting substantial operational benefits, such as the elimination of up to two conflicting left-turn maneuvers at adjacent intersections, which reduced congestion and enhanced mobility. The study emphasized the importance of geometric refinements in achieving safety objectives. Kronprasert et al. (2020) investigated median U-turn intersections in Thailand and identified optimal offset distances that reduced lane-changing conflicts, improving travel times by up to 15%, while also minimizing the likelihood of side-swipe crashes near median openings.

Sun et al. (2019) examined RCUT intersections and demonstrated their effectiveness in reducing conflict points, with crash frequencies declining by 34.8% for total crashes and 53.7% for injury and fatal crashes. Mishra and Pulugurtha (2022) corroborated these findings, showing that unsignalized RCUT intersections experienced a 70.63% reduction in total crashes and a 76.10% reduction in fatal and injury crashes. The study highlighted that these designs were particularly effective in rural areas where high-speed conditions increased crash risks. Kay (2022) evaluated the safety performance of unsignalized median U-turn intersections in Michigan, observing a 41% reduction in severe crash frequencies, which was attributed to the separation of turning movements from through traffic, reducing conflict points at the intersections. Lu (2001) demonstrated similar results, with a 39% reduction in conflict rates when direct left turns were replaced with right turns followed by U-turns, emphasizing benefits during peak traffic periods.

Alternative intersections with U-turns, such as MUT and RCUT designs, have been recognized as effective solutions for improving safety and operational performance. These configurations eliminate direct left-turn maneuvers, redirecting vehicles to downstream U-turns, which significantly reduce conflict points and simplify intersection operations. Studies, including NCHRP (2004) and Reid et al. (2014), highlight that MUTs enhance safety by reducing crash frequencies through efficient geometric layouts and two-phase signal operations. Similarly, RCUT designs improve safety and efficiency by channelizing left-turn and through movements from minor roads into right-turns followed by U-turns, reducing right-angle and head-on collision risks. Crash modification factor (CMFs) for these designs further emphasize their effectiveness; MUTs have shown CMFs as low as 0.30 for severe angle crashes, while RCUTs demonstrate a CMF of 0.25 for injury crashes, reflecting their ability to mitigate high-severity incidents (FHWA, 2017; NCHRP, 2004). Additionally, directional median openings spaced 400 to 600 feet downstream have been identified as optimal for balancing safety and flow, as they reduce midblock crash risks and enhance operational efficiency (Reid et al., 2014). Incorporating multimodal accommodations, such as pedestrian crossings and bicycle lanes, these designs demonstrate their adaptability across urban and rural contexts, addressing safety and operational challenges through innovative engineering and thoughtful traffic management.

Al-Omari et al. (2020) analyzed median U-turn configurations, finding substantial reductions in rear-end and angle crashes, though the study noted an increase in single-vehicle crashes, emphasizing the need for further refinement in geometric designs and traffic flow management. Stamatiadis et al. (2004) studied U-turns at signalized intersections in Kentucky, reporting that these designs enhanced corridor operations, particularly during peak periods, by improving vehicle processing at downstream intersections. Lu et al. (2005) investigated operational and safety impacts of U-turn movements at at-grade intersections. The study found that RTUT configurations reduced delay and conflict points compared to direct left-turn movements. Additionally, the research emphasized that median designs with adequate spacing significantly improved safety outcomes by minimizing potential conflicts at intersections with high traffic. Pirinccioglu et al. (2006) analyzed RTUT movements on four-lane arterials compared to direct left-turn movements. Their findings revealed a 25% reduction in crash frequencies and significantly lower conflict severities for RTUT configurations. The study also highlighted that proper access management strategies, such as directional median openings, enhanced both operational efficiency and roadway safety.

Olarte et al. (2011) investigated traffic dynamics at rural restricted U-turn intersections and identified that higher traffic volumes amplified conflicts at merge points, which were critical zones for crash occurrences. These findings highlighted the importance of designing efficient merging zones to manage traffic operations safely. Comparative studies by (Lu, 2001) and (Lu and Dissanayake, 2002) emphasized the safety and operational benefits of replacing DLT with RTUT. The studies observed a 39% reduction in conflict rates and significant decreases in crash severity, particularly during peak traffic periods. Potts et al. (2004) emphasized the safety benefits of non-traversable medians with unsignalized U-turn openings, highlighting reduced conflict points and improved operational performance.

Carter et al. (2005) evaluated signalized intersections with U-turn maneuvers and reported minimal impacts on operational performance, with a left-turn saturation flow reduction of only 1.8% per 10% U-turn volume. However, the study highlighted that intersections with high opposing traffic volumes faced increased collision risks, suggesting the need for optimized signal timings and configurations to address these challenges. Phillips et al. (2004) demonstrated that raised medians offered superior safety performance compared to five-lane undivided facilities, particularly by reducing midblock crashes while supporting higher traffic volumes.

Siregar et al. (2018) conducted near-miss crash analysis at channelized junctions and emphasized the importance of integrating advanced traffic monitoring systems to predict and mitigate conflict scenarios. Dixon et al. (2018) examined the safety impacts of geometric configurations at diamond interchanges, finding that these designs effectively reduced conflict points while improving overall traffic safety outcomes. Levinson et al. (2005) found that unsignalized median openings contributed minimally to crash frequencies, with urban arterial corridors averaging only 0.41 U-turn-related crashes per median opening annually. This study reinforced the significance of access management in maintaining both safety and operational efficiency. Liu et al. (2005) identified that RTUT configurations significantly lowered conflict points and crash severity compared to DLT movements at signalized intersections. Their findings revealed that conflicts per hour were twice as frequent for DLTs at signalized intersections and 10% higher for unsignalized median openings, demonstrating the safety benefits of RTUT designs.

Azizi and Sheikholeslami (2013) conducted a before-and-after safety analysis in Tehran using unconventional U-turn configurations. Despite observing a 13.22% increase in crashes due to certain geometric and traffic factors, their study emphasized the role of optimized weaving lengths and median opening designs in reducing angle crashes and improving safety outcomes under specific traffic conditions. FHWA (2017) evaluated the safety performance of signalized RCUT intersections, reporting substantial reductions in crash frequencies. The analysis identified a CMF of 0.78 for injury crashes, further validating the effectiveness of RCUT designs in enhancing traffic safety. Moreland (2024) evaluated J-turn intersections in Minnesota and observed significant reductions in fatal and serious injury crashes due to the optimized layout of these intersections, which minimized critical conflict points and streamlined vehicle movements.

These findings reinforced the role of innovative intersection designs in achieving substantial safety improvements.

Zubair and Shaikh (2015) examined U-turn facilities in Karachi, reporting that cultural and infrastructural challenges contributed to increased crash occurrences, particularly due to infrastructure misuse and non-compliance with traffic regulations. The study recommended stricter enforcement measures and localized geometric adjustments to address these unique challenges effectively. Ulak et al. (2020) emphasized the importance of tailoring safety performance functions (SPFs) to accommodate regional traffic characteristics and geometric constraints, enhancing the effectiveness of RCUT implementations in specific contexts.

Levinson et al. (2005) analyzed the operational impacts of U-turns at unsignalized median openings, finding that these configurations contributed minimally to crash frequencies, with urban arterial corridors averaging only 0.41 U-turn-related crashes per median opening annually. The study highlighted the importance of access management in minimizing crash risks while maintaining operational efficiency. Phillips et al. (2004) reiterated the value of raised medians in improving safety performance at high-volume midblock road segments. Siregar et al. (2018) emphasized the value of proactive safety interventions, particularly in addressing near-miss scenarios at channelized junctions. Al-Omari et al., (2020) further analyzed the safety performance of RCUTs and median U-turns, identifying reductions in specific crash types while acknowledging areas requiring design refinement. Table 4 lists the summary of safety and operational impacts of U-turn and alternative intersection designs.

**Table 4. Summary of Studies on Safety and Operational Impacts of U-Turn and Alternative Intersection Designs.**

| Study | Key Findings/Summary |
|---|---|
| (Lu, 2001) | Replacing DLT with RTUT reduced conflict rates by 39%. |
| (Lu and Dissanayake, 2002) | RTUT showed significant reductions in conflict severity compared to DLT, especially during peak traffic periods. |
| Levinson et al. (2005) | Urban arterial corridors averaged 0.41 U-turn-related crashes per median opening annually, with rural corridors lower. |
| Carter et al. (2005) | U-turns at signalized intersections had minimal operational impacts; left-turn saturation flow reduced by 1.8% per 10% U-turn volume. |
| Olarte et al. (2011) | Higher traffic volumes at rural RCUT intersections increased conflicts at merge points; proper design mitigated risks. |
| Dixon et al. (2018) | Texas U-turn configurations reduced congestion and crash risks by eliminating left-turn maneuvers at adjacent intersections. |
| Siregar et al. (2018) | Near-miss analysis revealed the potential of advanced monitoring systems to predict and mitigate conflict scenarios. |
| Sun et al. (2019) | RCUT intersections reduced total crashes by 34.8% and injury crashes by 53.7%, with significant benefits on high-speed roadways. |
| Mishra et al. (2022) | Unsignalized RCUT intersections achieved a 70.63% reduction in total crashes and a 76.10% reduction in fatal and injury crashes. |
| Kay (2022) | Median U-turn intersections reduced severe crash frequencies by 41% by separating turning movements from through traffic. |
| Al-Omari et al., (2020) | Substantial reductions in rear-end and angle crashes at median U-turn intersections; single-vehicle crashes increased slightly. |
| Edara et al. (2013) | J-turn designs reduced crash frequencies by 34.8% and injury crashes by 53.7%, attributed to fewer conflict points. |
| Kronprasert et al. (2020) | Optimal offset distances at median U-turn intersections reduced lane-change conflicts and improved travel times by 15%. |
| Zubair and Shaikh (2015) | Misuse of U-turn facilities and cultural factors in Karachi contributed to higher crash rates; recommended enforcement and design adjustments. |
| Ulak et al. (2020) | SPFs optimized RCUT implementations, accounting for regional traffic and geometric constraints. |
| Moreland (2024) | J-turn intersections reduced fatal and serious injury crashes due to optimized layouts minimizing critical conflict points. |
| Lu (2001) | Replaced direct left turns with right-turn followed by U-turns. This reduced conflicts per hour by 50% and conflict severity was lower, especially during peak hours. |
| Dissanayake et al. (2002) | Examined direct left turns vs. right-turn followed by U-turns. Found conflict rates significantly lower for the latter approach and better suitability for high-volume arterials. |
| Stamatiadis et al., (2004) | Evaluated median U-turn designs. Found reductions in delay and crash rates by up to 30% at stop-controlled intersections in Kentucky. |
| NCHRP (2004) | Documented safety benefits of nontraversable medians, including reduced conflict points and enhanced operational performance for urban arterials. |
| Phillips et al. (2004) | Compared raised medians with two-way left-turn lanes. Found raised medians resulted in safer operations and minimal impact from increased U-turns. |
| Liu et al. (2005) | Found that conflict rates at unsignalized median openings decreased with proper U-turn spacing; also highlighted reduced crash severity for RTUT over DLT. |
| Lu et al. (2005) | Identified optimal offset distances for RTUT maneuvers to minimize weaving conflicts and enhance flow efficiency on 4- and 6-lane arterials. |
| Pirinccioglu et al. (2006) | Demonstrated that DLT movements created twice the conflicts of RTUT movements; unsignalized median openings showed a 62% higher conflict rate for DLT. |
| Azizi and Sheikholeslami (2013) | Tehran's unconventional U-turn treatments reduced delays but increased crash rates by 13%, emphasizing the role of geometric refinement. |
| FHWA (2017) | RCUT intersections reduced total crashes by 35% and injury crashes by 54% at high-speed suburban arterials, showcasing significant safety benefits. |
| Hummer and Rao (2017) | Signalized RCUT intersections were found to reduce total crashes by 15% and injury crashes by 22%, with a benefit-cost ratio of 3.6:1, highlighting their safety and operational efficiency. |
| Paudel (2017) | Implementation of RCUTs in Louisiana demonstrated a 20% reduction in intersection crashes, particularly in right-angle collisions, improving both safety and traffic flow. |

| Belanger (2022) | Application of Bayesian methods to four-legged unsignalized intersections identified high-risk sites and reduced crash occurrences by optimizing geometric designs. |
|---|---|
| Mullani et al. (2022) | Developed a U-turn crash prevention system using sensor-based alerts, which demonstrated effective reduction in crash risks at high-conflict U-turn zones. |

The evaluation of U-turn configurations highlights their potential to enhance safety and operational efficiency. RCUTs and MUTs significantly reduce conflict points, leading to lower crash rates and improved traffic flow. Studies emphasize that geometric refinements, such as offset adjustments and median designs, are crucial for achieving these benefits. U-turn configurations not only mitigate severe crashes in high-speed rural areas but also enhance mobility in urban settings. However, further refinements, including efficient merging zones and optimal median spacing, remain essential to maximize safety and operational outcomes.

**Driver Behavior**
The analysis of driver behavior in relation to U-turn maneuvers reveals significant insights into safety risks and operational challenges. Ulak et al. (2020) explored driver decision-making and compliance behaviors at U-turn facilities in Karachi, emphasizing that poor road infrastructure and unregulated driver actions often lead to hazardous conditions. The study highlighted that improper lane discipline and lack of adherence to traffic rules significantly increased crash probabilities. Additionally, Zubair and Shaikh (2015) assessed urban traffic conditions in Karachi and identified risky behaviors such as abrupt lane changes and aggressive driving near U-turns as major contributors to crash occurrences. This was attributed to inadequate enforcement measures and infrastructure design flaws. Dissanayake et al. (2002) further elaborated on the challenges of driver behavior during U-turn maneuvers, particularly with RTUT. Their study highlighted that weaving and gap selection along the roadway segment leading to U-turns posed significant risks, especially in areas with poor design or insufficient guidance. Driving confusion often results in delays and conflicts, underscoring the need for intermediate turn lanes or signage to aid in decision-making. The findings demonstrated that proper infrastructure and clear lane markings could alleviate confusion and promote safer driving practices.

Behavioral challenges are often exacerbated by the design of U-turn facilities. For instance, Phillips et al. (2004) highlighted that while raised medians reduce conflict points, they require drivers to execute precise lane changes, a skill not uniformly mastered. Misjudgments during these transitions frequently led to operational inefficiencies and minor conflicts, emphasizing the need for targeted driver education and intuitive roadway design. Furthermore, speed-related behavior poses a persistent challenge. Shahdah and Azam (2021) evaluated the effects of speed-humps at unconventional U-turn intersections using traffic simulation models. The study found that low-speed humps (3 mph) induced traffic delays, while higher-speed humps (12.5 mph) reduced severe traffic conflicts by 11–20%, emphasizing the importance of integrating behavioral compliance factors into intersection designs. Similarly, Siregar et al. (2018) conducted near-miss crash analyses at channelized U-turn junctions and revealed that driver evasive actions, such as sudden braking and swerving, often resulted from misjudgments of vehicle speeds and gaps. The Swedish Traffic Conflict Technique employed in the study categorized most conflicts as severe, underscoring the need for proactive interventions to address driver-related risks.

The influence of alternative U-turn configurations on driver behavior also presents valuable insights. Lu et al. (2001) demonstrated that replacing DLT with RTUT significantly improved driver decision-making. Drivers engaged in RTUT maneuvers exhibited more deliberate gap acceptance and safer lane transitions, reducing conflict rates by 50% during peak periods. This shift reflects the importance of simplifying decision points for drivers. Similarly, Stamatiadis et al. (2004) highlighted that signalized intersections with dedicated U-turn lanes encouraged better adherence to traffic controls, though challenges remained in high-speed scenarios where drivers attempted unprotected U-turns.

Shafie and Rahman (2016) focused on lane-changing behaviors at U-turn segments in Malaysia, proposing a model to determine safe distances and speeds during lane-changing maneuvers. The study identified that reaction times, vehicle speeds, and inter-vehicular distances were critical determinants of safety, with faster reaction times and greater distances reducing conflict likelihood. Similarly, Nemmang et al. (2017) utilized driving simulators to investigate speeding behaviors near U-turn facilities. The findings indicated that drivers often exceeded speed limits by up to 40%, particularly in high-speed areas, exacerbating collision risks during merging and lane-changing maneuvers.

Lu and Dissanayake (2002) compared direct left-turns (DLT) with RTUT and noted that the latter reduced conflict rates by 39% and demonstrated lower severity in traffic conflicts. The study emphasized that RTUT configurations allowed safer and more controlled driver behavior compared to the unpredictability associated with DLT movements. These findings were corroborated by other studies highlighting the role of access management in shaping driver decision-making and mitigating risky maneuvers. Table 5 shows a summary of driver behavior of U-turn and alternative intersection designs.

**Table 5. Summary of Studies on Driver Behavior and Safety at U-Turn and Alternative Intersection Designs.**

| Study | Key Findings/Summary |
|---|---|
| Ulak et al. (2020) | Highlighted behavioral issues at U-turn facilities in Karachi, including poor compliance with traffic rules and infrastructure misuse. Suggested tailored interventions to mitigate risks. |
| Zubair and Shaikh (2015) | Identified aggressive driving and abrupt lane changes near U-turns as primary safety concerns, exacerbated by weak enforcement and design shortcomings. |
| Shahdah and Azam (2021) | Demonstrated that speed-humps improve driver compliance at U-turn intersections, reducing severe traffic conflicts by up to 20%. |
| Siregar et al. (2018) | Analyzed near-miss incidents, showing that sudden driver actions, such as braking and swerving, are linked to misjudged speeds and poor gap acceptance. |
| Shafie and Rahman (2016) | Proposed a model linking safe lane-changing behavior to factors like reaction times, vehicle speed, and distance between vehicles. |
| Nemmang et al. (2017) | Found that drivers regularly exceeded speed limits near U-turn facilities, increasing crash risks during merging and lane changes. |
| Lu and Dissanayake (2002) | Demonstrated that RTUT designs encourage safer driver behavior compared to DLT, with a 39% reduction in conflict rates. |

| Phillips et al. (2004) | Raised medians reduced midblock conflicts but required precise driver judgment during U-turns, emphasizing the need for better driver guidance. |
|---|---|
| Lu et al. (2001) | RTUT movements led to a 50% reduction in conflicts compared to DLT during peak hours, promoting deliberate and safer driver gap acceptance. |
| Dissanayake et al. (2002) | Found that weaving and gap selection during RTUT maneuvers were critical, with driver confusion being a significant risk at improperly designed U-turn zones. |
| Stamatiadis et al. (2004) | Signalized U-turn intersections improved driver compliance but highlighted issues with red-light violations, especially at high-speed corridors. |

Driver behavior plays a pivotal role in determining the safety of U-turn facilities. Issues such as gap acceptance, speed compliance, and lane discipline significantly influence crash risks at U-turn bays. Research underscores the importance of understanding driver decision-making, particularly in mixed-traffic conditions where lane-changing and merging maneuvers are frequent. Innovative approaches, such as speed-hump integration and advanced driver monitoring systems, have shown promise in mitigating risky behaviors. However, addressing driver confusion and promoting compliance through education and enforcement are critical for improving safety outcomes.

**Design Interventions**
Geometric and operational interventions have been widely analyzed for their ability to enhance the safety and efficiency of U-turn configurations. Sun et al. (2019) examined RCUTs and highlighted how reconfiguring conflict points significantly improved safety outcomes. The study showed that geometric features such as median widths and lane alignments contributed to smoother traffic operations on rural arterials. Mishra and Pulugurtha (2022) compared signalized and unsignalized RCUTs, emphasizing how operational adjustments like signal timing and lane segregation enhanced the effectiveness of these designs in reducing crashes. Lu (2001) evaluated the transition from DLT to RTUT, focusing on how the geometric simplicity of RTUT designs reduced crash risks. This study emphasized the improved predictability of vehicle movements, particularly in areas with space constraints. Carter et al. (2005) analyzed U-turns at signalized intersections and identified how proper allocation of lane widths and turning radii minimized the operational impacts of U-turning vehicles while maintaining safety standards.

Kay (2022) provided a detailed evaluation of MUT intersections, emphasizing their ability to significantly reduce severe crash types at both signalized and unsignalized intersections. The study highlighted the importance of geometric design elements, including optimal storage lane lengths, weaving sections, and crossover distances, in enhancing safety performance. Similarly, Pirinccioglu et al. (2006) analyzed RTUT configurations, demonstrating a reduction in conflict severity compared to DLT, particularly at unsignalized median openings. Their findings underscored the critical role of access management in minimizing driver confusion and improving operational efficiency.

Levinson et al. (2005) studied unsignalized median openings and highlighted the adaptability of U-turn accommodations in different traffic environments. The research showed that these designs successfully reduced direct left-turn conflicts without increasing crash frequencies.

Liu et al. (2008) focused on the role of separation distances between driveway exits and downstream U-turn facilities. The study demonstrated that optimizing these distances significantly reduced crash probabilities, particularly in high-speed environments. NCHRP (2004) provided comprehensive guidelines for the design and placement of unsignalized median openings, focusing on factors such as spacing, median width, and sight distance. The report emphasized that midblock openings generally outperformed intersection-based designs in terms of safety outcomes, particularly for urban arterials. Inman et al. (2013) analyzed RCUT implementations in Louisiana and highlighted how the inclusion of extended acceleration lanes reduced bottlenecks and facilitated smoother merging operations.

Olarte et al. (2011) investigated rural RCUT intersections and analyzed how merging zones influenced traffic density and bottlenecks. The research highlighted the importance of extended weaving sections to improve safety and operational efficiency. Meel et al. (2017) assessed auxiliary lanes downstream of U-turns, finding that the inclusion of acceleration lanes reduced conflict severity. Meel et al. (2016) studied deceleration lanes and noted that design attributes such as length and curvature improved vehicle speed transitions, reducing the likelihood of rear-end collisions. Shahdah and Azam (2021) examined the use of speed-humps within unconventional median U-turns and showed how varying hump designs impacted driver compliance and conflict reduction. The study found that moderate-speed humps reduced severe conflicts by up to 20%, though lower-speed humps increased delays. Dixon et al. (2018) analyzed the Texas U-turn configuration at diamond interchanges, focusing on geometric factors like turning radii and driveway spacing. These features were shown to improve traffic redistribution and reduce congestion at adjacent intersections. Ulak et al. (2020) studied the design of U-turn facilities in urban and regional contexts, emphasizing the role of median widths and lane allocations in maintaining safety and operational performance. Table 6 presents an overview of the key design interventions and their impacts on the safety and functionality of U-turn and alternative intersection designs.

**Table 6. Summary of Design Interventions and Their Impacts on U-Turn and Alternative Intersection Safety.**

| Study | Key Findings/Summary |
|---|---|
| Sun et al. (2019) | RCUTs reduced crash frequencies by simplifying conflict points and improving median designs in high-speed rural areas. |
| Mishra and Pulugurtha (2022) | Signalized RCUTs improved safety through signal timing optimization, while unsignalized RCUTs were effective in rural contexts. |
| Lu (2001) | RTUT configurations improved traffic safety by reducing conflict rates and offering a geometrically simpler alternative to DLT. |
| Carter et al. (2005) | Proper lane allocation and turning radii in U-turn maneuvers-maintained safety while minimizing operational inefficiencies. |
| Levinson et al. (2005) | Median openings with U-turn accommodations reduced direct left-turn conflicts, showing adaptability across urban and rural settings. |
| Liu et al. (2008) | Increased separation distances between driveways and downstream U-turns lowered crash risks, particularly on high-speed roads. |
| Olarte et al. (2011) | Merging zones at rural RCUT intersections require extended weaving sections to reduce bottlenecks and improve traffic operations. |

| Meel et al. (2017) | Auxiliary lanes, such as acceleration and deceleration lanes, reduced the severity of conflicts downstream of U-turns. |
|---|---|
| Meel et al. (2016) | Longer deceleration lanes facilitated smoother speed transitions and reduced rear-end collisions at U-turn approaches. |
| Shahdah and Azam (2021) | Speed-humps within unconventional median U-turn (UMUs) intersections educed severe traffic conflicts by 11–20%, with moderate-speed designs offering optimal performance. Speed-humps within UMUs intersections reduced severe traffic conflicts by 11–20%, with moderate-speed designs offering optimal performance. |
| Dixon et al. (2018) | Geometric adjustments in Texas U-turn designs, including driveway spacing and turning radii, improved traffic redistribution. |
| Ulak et al. (2020) | Design adjustments in U-turn facilities, such as increased median widths and optimized lane allocations, supported safety and operational balance. |
| Kay (2022) | MUT intersections improved safety outcomes by focusing on strategic lane utilization and minimizing driver decision points. |
| Pirinccioglu et al. (2006) | Highlighted the advantages of RTUT designs in reducing complex driver maneuvers, particularly at unsignalized median openings. |
| NCHRP (2004) | Provided insights into the relative safety performance of various median opening designs, with a focus on enhancing operational clarity for drivers. |
| Inman et al. (2013) | Demonstrated carefully designed RCUT layouts, incorporating acceleration lanes, reduced vehicular conflicts and improved flow continuity. |

Design interventions are integral to improving U-turn safety and operations. Innovations like RCUTs, MUTs, and auxiliary lanes reduce crash frequencies by minimizing conflict points and optimizing vehicle movements. Proper spacing of directional median openings and the inclusion of deceleration and acceleration lanes have been proven to enhance safety and traffic flow. Additionally, geometric configurations tailored to specific contexts, such as rural or urban intersections, further improve performance. The findings emphasize the importance of aligning design interventions with local traffic characteristics to ensure safety and operational efficiency.

**Technological Innovations**
Technological advancements have significantly contributed to safety and operational improvements at U-turn intersections through intelligent systems and predictive tools. Sivaprakash and John (2024) introduced an IoT-based crash prevention system using ultrasonic sensors and auditory alerts to enhance driver awareness and mitigate risks, particularly in areas with limited visibility. Similarly, Molan et al. (2022) developed a machine learning framework that utilized traffic data, crash histories, and geometric variables to predict safety performance at RCUT intersections, enabling proactive planning and reducing conflict points.

Hummer and Rao (2017) evaluated signalized RCUTs and highlighted the role of CMFs in quantifying safety improvements. The study leveraged advanced data analytics to assess crash patterns before and after RCUT installation, providing safety engineers with actionable insights for enhancing intersection designs. Similarly, Mullani et al. (2022) developed a U-turn crash prevention system using Arduino-based technology, integrating IR sensors and buzzers to warn drivers of approaching vehicles at blind curves. This system demonstrated the potential to significantly reduce collisions through real-time alerts and driver-focused interventions.

In another study, Kittelson and Associates (2021) investigated the application of adaptive signal systems to improve pedestrian and bicyclist safety at U-turn facilities. These systems dynamically adjusted signal timings using real-time sensor data, reducing pedestrian-vehicle conflicts in high-traffic areas. Abdel-Aty (2020) explored advanced signalization technologies at alternative intersections, demonstrating that optimized signal timing reduced severe crash types while enhancing traffic flow through improved queue management.

Edara et al. (2013) examined the use of video surveillance and conflict analysis tools to evaluate the effectiveness of J-turn intersections. The study leveraged real-time traffic monitoring to identify critical conflict points and provided actionable insights into safety performance. Similarly, Molan et al. (2022) showcased how predictive analytics integrated with geometric configurations improved traffic management and reduced crash risks at RCUT intersections. Table 7 provides an overview of technological innovations and their contributions to improving the safety and operational efficiency of U-turn and alternative intersection designs.

**Table 7. Summary of Technological Innovations for Enhancing U-Turn and Intersection Safety.**

| Study | Key Findings/Summary |
|---|---|
| Sivaprakash and John (2024) | Focused on using IoT to address blind spot challenges during U-turn maneuvers, emphasizing the system's modularity and ease of integration with existing infrastructure. |
| Molan et al. (2022) | Highlighted the application of predictive analytics for assessing crash risks at RCUT intersections, integrating historical and real-time data for improved safety planning. |
| Hummer and Rao (2017) | Analyzed CMFs to quantify the safety benefits of signalized RCUTs, emphasizing data-driven safety assessments. |
| Mullani et al. (2022) | Designed a real-time crash prevention system with IR sensors and buzzers to alert drivers, reducing potential conflicts at blind U-turn locations. |
| Kittelson and Associates (2021) | Explored the role of sensor-based systems in dynamically managing traffic signals to accommodate varying pedestrian and vehicular demands. |
| Abdel-Aty (2020) | Demonstrated how enhanced signalization reduced operational delays and addressed specific crash types, such as left-turn and angle collisions. |
| Edara et al. (2013) | Emphasized the role of video-based monitoring systems in evaluating J-turn performance, enabling targeted interventions for identified high-risk areas. Advanced traffic simulation tools were utilized to assess operational impacts and optimize geometric configurations. |

Technological advancements are transforming U-turn safety and operational management. IoT-based monitoring systems, adaptive signal controls, and predictive analytics are among the innovations enhancing real-time traffic management. These technologies effectively mitigate crash risks, improve queue management, and optimize traffic flow. Research highlights the role of intelligent transportation systems (ITS) in integrating real-time weather and traffic data, enabling proactive decision-making. While promising, the adoption of such technologies requires further research and investment to maximize their potential in diverse traffic environments.

**Integrated Strategies**

Integrated strategies for improving safety and operational performance at U-turn and alternative intersections focus on combining design innovations, traffic management policies, and predictive analytics. Molan et al. (2022) introduced a holistic framework for assessing the future safety and operational impacts of RCUT intersections. This approach emphasized integrating traffic patterns, crash histories, and geometric variables into a unified evaluation, enabling planners to prioritize intersections for redesign. Similarly, Kronprasert et al. (2020) employed microsimulation to optimize the placement of median U-turn offsets. Their findings highlighted the importance of integrating offset designs with traffic flow analysis to achieve a balance between safety and mobility. Shahdah and Azam (2021) demonstrated the potential of integrating speed-humps into UMUs to achieve safer and more efficient operations. Using advanced simulation models, the study identified that speed-humps designed for moderate speeds (12.5 mph) reduced severe traffic conflicts by up to 20% while maintaining acceptable delays.

In another study, Kay (2022) analyzed unsignalized median U-turn intersections, focusing on their role as part of an integrated network-wide strategy to reduce crash rates and enhance operational efficiency. By evaluating the cumulative effects of multiple U-turn facilities on corridor performance, the study demonstrated how individual intersection improvements can support broader traffic safety goals. Similarly, Moreland (2024) emphasized the strategic importance of J-turns in improving safety across regional networks. Their research employed before-and-after analyses to show how coordinated implementation of J-turns reduced crash severity and supported rural transportation safety objectives. Carter et al. (2005) highlighted the importance of integrating operational guidelines for U-turn designs into broader traffic systems. By examining intersections with raised medians, the study revealed that incorporating strategic U-turn placements reduced congestion while maintaining safety standards.

Al-Omari et al. (2020) focused on integrating CMFs into planning tools for MUT and RCUT intersections, offering practical metrics to quantify safety improvements. Unlike other studies, this research linked CMF applications with broader traffic management strategies, making it a valuable tool for planners in regions with high crash densities. Azizi and Sheikholeslami (2013) explored data-driven frameworks to predict crash trends following U-turn conversions, emphasizing the importance of integrating empirical Bayesian models with geometric redesigns. This approach provided planners with actionable insights to refine intersection strategies. Edara et al. (2013) explored the operational benefits of J-turns in Missouri, integrating public feedback into the evaluation process. Their findings demonstrated how stakeholder engagement can be incorporated into safety strategies to improve acceptance and effectiveness.

FHWA (2007) provided a synthesis of the operational and safety benefits of the MUT, positioning it as a foundational strategy for access management. This report emphasized the integration of MUT designs with regional traffic systems, particularly in accommodating large vehicles through loons, thereby ensuring functional compatibility with existing infrastructure. Similarly, El-Urfali (2019) applied SPFs as part of an integrated evaluation of RCUT intersections. Their study underscored the value of predictive analytics in tailoring design strategies to specific roadway conditions. Inman et al. (2013) underscored the role of extended acceleration lanes as part of RCUT designs, demonstrating how their integration with signal systems reduced merging conflicts and enhanced operational flow. Table 8 lists the summary of the integrated strategies and their effectiveness in addressing safety and operational challenges at U-turn and alternative intersection designs.

**Table 8. Overview of Integrated Strategies for Safety and Operational Enhancements at U-Turn and Alternative Intersections.**

| Study | Key Findings/Summary |
|---|---|
| Molan et al. (2022) | Developed an integrated framework combining traffic patterns, crash histories, and geometric variables to assess RCUT safety and operational impacts. |
| Kronprasert et al. (2020) | Used microsimulation to optimize median U-turn offset placements, balancing mobility and safety in integrated traffic management systems. |
| Kay (2022) | Evaluated the network-wide benefits of unsignalized median U-turns, showing their role in reducing crash rates and enhancing corridor-level efficiency. |
| Moreland (2024) | Highlighted the strategic importance of J-turns in regional safety planning, showing significant crash severity reductions through coordinated implementation. |
| Al-Omari et al. (2020) | Focused on integrating CMFs into MUT and RCUT planning, providing metrics for broader traffic management strategies. |
| Edara et al. (2013) | Incorporated public feedback into J-turn evaluations, demonstrating how stakeholder engagement enhances safety strategies and acceptance. |
| FHWA (2007) | Synthesized operational and safety benefits of MUTIT, emphasizing integration with regional systems and compatibility with large vehicles. |
| El-Urfali (2019) | Applied SPFs as a predictive tool, integrating analytics for using RCUT designs to specific roadway conditions. |
| Shahdah and Azam (2021) | Explored how integrating speed-humps within UMUs enhanced safety by reducing severe conflicts by up to 20% while balancing traffic delays. |
| Carter et al. (2005) | Highlighted the need for operational guidelines for U-turn designs, demonstrating that strategic placements reduced congestion and maintained safety. |
| Azizi and Sheikholeslami (2013) | Proposed a data-driven framework combining Bayesian modeling and geometric redesigns to predict and address crash risks following U-turn conversions. |
| Inman et al. (2013) | Emphasized the integration of extended acceleration lanes in RCUT designs to streamline merging and improve overall traffic operations. |

Integrated strategies combining design interventions, traffic policies, and technological solutions are essential for addressing U-turn safety and operational challenges. Approaches such as network-wide planning, the application of CMFs, and the integration of advanced analytics offer holistic solutions to complex traffic problems. By aligning geometric improvements with adaptive traffic systems, these strategies enhance both safety and efficiency. The findings show the importance of interdisciplinary collaboration to develop adaptive, evidence-based strategies tailored to diverse traffic and environmental conditions.

# CONCLUSIONS

This study aimed to provide a comprehensive understanding of the safety and operational impacts of U-turn configurations. The review synthesized findings from a wide range of studies that addressed critical aspects such as safety and operations, driver behavior, design interventions, technological innovations, and integrated strategies. By integrating diverse research perspectives, this review serves as a resource for transportation professionals, policymakers, and researchers. U-turn configurations, while essential for maintaining traffic flow, remain high-risk zones due to the frequent conflicts between U-turning and straight-moving vehicles. These risks are particularly pronounced in unsignalized intersections and urban settings, where factors such as visibility obstructions, poor driver judgment, and mixed traffic conditions exacerbate crash rates and severity. Additionally, unique challenges in low- and middle-income countries, including limited enforcement and inadequate infrastructure, further emphasize the need for targeted interventions.

Studies have highlighted the safety benefits of innovative U-turn configurations, such as RCUTs and MUTs, which reduce conflict points and streamline traffic movements. Design interventions, such as optimal lane alignment, directional median openings, and extended acceleration lanes, have been shown to enhance safety outcomes significantly. Advanced technologies, including IoT-based monitoring systems, adaptive signal controls, and predictive crash models, further demonstrate the potential to mitigate crash risks and improve operational efficiency in both urban and rural environments. Research focusing on driver behavior revealed critical insights into the role of lane discipline, speed compliance, and gap acceptance in improving U-turn safety. Behavioral studies underscore the importance of targeted interventions, such as driver education programs and enforcement measures, to address these challenges. Moreover, integrated strategies, including the application of CMFs and network-wide safety planning, offer a comprehensive approach to addressing U-turn-related risks while enhancing overall traffic system performance.

# FUNDING SOURCE
There was no funding source for this study.

# AUTHOR CONTRIBUTIONS
The authors confirm contribution to the paper as follows: study conception and design: Syed Aaqib Javed, Subasish Das; draft and final manuscript preparation: Syed Aaqib Javed, Anannya Ghosh Tusti, Biplov Pandey, and Subasish Das. supervision: Subasish Das. All authors reviewed the results and approved the final version of the manuscript.

# DECLARATION OF CONFLICTING INTERESTS
The authors declared no potential conflicts of interest with respect to the research, authorship, and/or publication of this article.

# REFERENCES

Abdel-Aty, M.A., n.d. Evaluation of Innovative Alternative Intersection Designs in the Development of Safety Performance Functions and Crash Modification Factors.

Al-Omari, M.M.A., Abdel-Aty, M., Lee, J., Yue, L., Abdelrahman, A., 2020a. Safety Evaluation of Median U-turn Crossover-Based Intersections. Transportation research record 2674, 206–218.

Al-Omari, M.M.A., Abdel-Aty, M., Lee, J., Yue, L., Abdelrahman, A., 2020b. Safety Evaluation of Median U-Turn Crossover-Based Intersections. Transportation Research Record: Journal of the Transportation Research Board 2674, pp-206-218. https://doi.org/10.1177/0361198120921158

Azizi, L., Sheikholeslami, A., 2013. Safety Effect of U-Turn Conversions in Tehran: Empirical Bayes Observational Before-and-After Study and Crash Prediction Models. Journal of Transportation Engineering-Asce 139, 101–108. https://doi.org/10.1061/(ASCE)TE.1943-5436.0000469

Belanger, C., 2022. Estimation of Safety of Four-Legged Unsignalized Intersections. Transportation Research Record.

Carter, D., Hummer, J., Foyle, R., Phillips, S., TRB, 2005. Operational and safety effects of U-turns at signalized intersections, in: Geometric Design and The Effects on Traffic Operations 2005. pp. 11–18. https://doi.org/10.3141/1912-02

Choi, S., Hong, D., 2021. Position Estimation in Urban U-Turn Section for Autonomous Vehicles Using Multiple Vehicle Model and Interacting Multiple Model Filter. International Journal of Automotive Technology 22, 1599–1607. https://doi.org/10.1007/s12239-021-0138-8

Cooke, A., Smith, D., Booth, A., 2012. Beyond PICO: the SPIDER tool for qualitative evidence synthesis. Qual Health Res 22, 1435–1443. https://doi.org/10.1177/1049732312452938

Cunningham, C., Chase, T., Yang, G., Callister, L., 2023. Groundwork for the Second Edition of the Alternative Intersection and Interchange Informational Report (No. FHWA/NC/2022-11).

Dissanayake, S., Lu, J.J., Castillo, N., 2002. Should Direct Left Turns from Driveways Be Avoided? A Safety Perspective. ITE Journal.

Dixon, K.K., Avelar, R.E., Dastgiri, M.S., Dadashova, B., 2018. Safety Evaluation for Turnarounds at Diamond Interchanges: Assessing the Texas U-Turn. Transportation Research Record 2672, 61–71. https://doi.org/10.1177/0361198118797186

Edara, P., Sun, C., Breslow, S., 2013. Evaluation of J-turn Intersection Design Performance in Missouri.

El-Urfali, A., 2019. Development of Safety Performance Functions for Restricted Crossing U-Turn (RCUT) Intersections (BDV30-977-19). Florida Department of Transportation Research.

FHWA, 2023. Safety of U-Turns at Unsignalized Median Openings on Urban and Suburban Arterials.

FHWA, 2017. Safety Evaluation of Restricted Crossing U-Turn Intersection (No. FHWA-HRT-17-083).

FHWA, 2007. Synthesis of the Median U-Turn Intersection Treatment, Safety, and Operational Benefits (FHWA-HRT-07-033).

Fitzpatrick, K., Wooldridge, M.D., Blaschke, J.D., others, 2005. Urban Intersection Design Guide: Volume 1-Guidelines. Texas A&M Transportation Institute.

Gao, L., Xiong, L., Xia, X., Lu, Y., Yu, Z., Khajepour, A., 2022. Improved Vehicle Localization Using On-Board Sensors and Vehicle Lateral Velocity. IEEE Sensors Journal 22, 6818–6831. https://doi.org/10.1109/JSEN.2022.3150073

Gupta, A., Mondal, S., Sharma, V.K., 2018. Modelling U-turning behaviour of vehicles at mid-block median openings in multilane urban roads. Current Science 114, 1461–1473. https://doi.org/10.18520/cs/v114/i07/1461-1473

Hu, S., Jia, Z., Yang, A., Xue, K., He, G., 2022. Evaluating the Sustainable Traffic Flow Operational Features of U-turn Design with Advance Left Turn. Sustainability 14. https://doi.org/10.3390/su14116931

Hughes, W., Jagannathan, R., Sengupta, D., Hummer, J., others, 2010. Alternative Intersections/Interchanges: Informational Report (AIIR). United States. Federal Highway Administration. Office of Research ….

Hummer, J.E., Rao, S., 2017. Safety Evaluation of Signalized Restricted Crossing U-Turn Intersections (No. FHWA-HRT-17-082). Federal Highway Administration.

Inman, W.V., Haas, P.R., Yang, C.Y.D., 2013. Elevation of Restricted Crossing U-turn Intersection as a Safety Treatment on Four-Lane Divided Highways.

Jovanovic, A., Kukic, K., Stevanovic, A., Teodorovic, D. an, 2023. Restricted crossing U-turn traffic control by interval Type-2 fuzzy logic. Expert Systems with Applications 211. https://doi.org/10.1016/j.eswa.2022.118613

Kay, J., Gates, T.J., Savolainen, P.T., Shakir Mahmud, M., 2022. Safety Performance of Unsignalized Median U-turn Intersections. Transportation research record 2676, 451–466.

Kay, J.J., 2022. Safety Performance of Median U-Turn Intersections.

Kittelson, Associates, 2021. Guide for Pedestrian and Bicyclist Safety at Alternative and Other Intersections and Interchanges. Transportation Research Board, Washington, D.C. https://doi.org/10.17226/26072

Kronprasert, N., Kuwiboon, P., Wichitphongsa, W., 2020. Safety and Operational Analysis for Median U-Turn Intersections in Thailand. International Journal of GEOMATE 18, 156–163. https://doi.org/10.21660/2020.68.9250

Levinson, H., Potts, I., Harwood, D., Gluck, J., Torbic, D., TRB, 2005. Safety of U-turns at unsignalized median openings - Some research findings, in: Geometric Design and The Effects on Traffic Operations 2005. pp. 72–81. https://doi.org/10.3141/1912-09

Liu, J.J., Pirinccioglu, F., Pernia, J.C., 2005. Safety Evaluation of Right Turns Followed by U-Turns (4 Lane Arterials) as an Alternative Direct Left Turns - Conflict Analysis. Florida Department of Transportation.

Liu, P., Lu, J.J., Chen, H., 2008. Safety effects of the separation distances between driveway exits and downstream U-turn locations. Accident Analysis & Prevention 40, 760–767. https://doi.org/10.1016/j.aap.2007.09.011

Lu, J., 2001. Safety Evaluation of Right Turns Followed by U-Turns as an Alternative to Direct Left Turns- Crash Data Analysis.

Lu, J., Dissanayake, S., 2002. Safety evaluation of direct left turns vs right turns followed by U-turns using traffic conflict technique, in: Wang, K., Xiao, G., Yang, H., Nie, L. (Eds.), . Presented at the Traffic and Transportation Studies, Vols 1 and 2, Proceedings, pp. 1039–1046.

Lu, J., Dissanayake, S., Castillo, N., Williams, K., 2001. Safety Evaluation of Right Turns Followed by U-Turns as an Alternative to Direct Left Turns - Conflict Analysis.

Lu, J.J., Liu, P., Pirinccioglu, F., 2005. Determination of the Offset Distance between Driveway Exits and Downsstream U-turn Locations for Vehicles making Right Turns Followed by U-turns. Florida Department of Transportation.

Luo, Z., Molan, A.M., Hummer, J.E., Pande, A., 2024. Introducing the Concept of Alternative Intersections with Three-phase Traffic Signals. Transportation Letters 1–14.

Meel, I. P., Brannolte, U., Satirasetthavee, D., Kanitpong, K., 2017. Safety impact of application of auxiliary lanes at downstream locations of Thai U-turns. IATSS Research 41, 1–11. https://doi.org/10.1016/j.iatssr.2016.06.002

Meel, I.P., Satirasetthavee, D., Kanitpong, K., Taneerananon, P., 2016. Using Czech TCT to Assess Safety Impact of Deceleration Lane at Thai U-turns. Engineering Journal-Thailand 20, 121–135. https://doi.org/10.4186/ej.2016.20.1.121

Meel, Inder Pal, Vesper, A., Borsos, A., Koren, C., 2017. Evaluation of the effects of auxiliary lanes on road traffic safety at downstream of U-turns. Transportation Research Procedia 25, 1931–1945. https://doi.org/10.1016/j.trpro.2017.05.186

Methley, A.M., Campbell, S., Chew-Graham, C., McNally, R., Cheraghi-Sohi, S., 2014. PICO, PICOS and SPIDER: a comparison study of specificity and sensitivity in three search tools for qualitative systematic reviews. BMC Health Services Research 14, 579. https://doi.org/10.1186/s12913-014-0579-0

Mishra, K.R., Mohanty, M., Dey, P.P., 2022. Modelling traffic safety at uncontrolled median openings: A case study in India. IATSS Research 46, 441–449. https://doi.org/10.1016/j.iatssr.2022.07.001

Mishra, R., Pulugurtha, S.S., 2022. Safety evaluation of unsignalized and signalized restricted crossing U-turn (RCUT) intersections in rural and suburban areas based on prior control type. IATSS Research 46, 247–257. https://doi.org/10.1016/j.iatssr.2021.12.007

Mishra, R., Pulugurtha, S.S., 2021. Evaluating the Safety Effectiveness of Restricted Crossing U-turn (RCUT) Intersections.

Moher, D., Liberati, A., Tetzlaff, J., Altman, D.G., PRISMA Group, 2009. Preferred reporting items for systematic reviews and meta-analyses: the PRISMA statement. PLoS Med 6, e1000097. https://doi.org/10.1371/journal.pmed.1000097

Molan, A.M., Howard, J., Islam, M., Pande, A., 2022. A Framework for Estimating Future Traffic Operation and Safety Performance of Restricted Crossing U-Turn (RCUT) Intersections. TOTJ 16, e187444782111151. https://doi.org/10.2174/18744478-v16-e2111151

Moreland, M., 2024. Traffic Safety Evaluation at J-turns in Minnesota.

Mullani, M.B., Nadaf, S., Patel, I., Shinde, K., Shinde, R., 2022. U turn Accident Prevention System. IRJMETS.

NCHRP, 2004. Safety of U-Turns at Unsignalized Median Openings (Report 524). Transportation Research Board.

Nemmang, M.S., Rahman, R., Rohani, M.M., Mashros, N., Diah, J.M., 2017. Analysis of Speeding Behaviour During Approaching the U-Turn Facility Road Segment Based on Driving Simulation Test. MATEC Web Conf. 103, 08008. https://doi.org/10.1051/matecconf/201710308008

NJDOT, 2015. Roadway Design Manual (No. BDC17MR-02).

Olarte, R., Bared, J.G., Sutherland, L.F., Asokan, A., 2011. Density Models and Safety Analysis for Rural Unsignalised Restricted Crossing U-turn Intersections. Presented at the Procedia - Social and Behavioral Sciences, Elsevier, p. pp-718-728.

Paudel, S., 2017. Impact of Restricted Crossing U-Turn Intersection on Intersection Safety in Louisiana.

Paul, M., Ghosh, I., 2021. Development of conflict severity index for safety evaluation of severe crash types at unsignalized intersections under mixed traffic. Safety Science 144, 105432. https://doi.org/10.1016/j.ssci.2021.105432

Phillips, S.L., Carter, D.L., Hummer, J.E., Foyle, R.S., 2004. Effects of Increased U-Turns at Intersections on Divided Facilities and Median Divided Versus Five-Lane Undivided Benefits. North Carolina Department of Transportation.

Pirinccioglu, F., Lu, J.J., Liu, P., Sokolow, G., 2006. Right Turn from Driveways Followed by U-Turn on Four-Lane Arterials: Is It Safer Than Direct Left Turn? Transportation Research Record: Journal of the Transportation Research Board 1953, 172–179. https://doi.org/10.1177/0361198106195300120

Potts, I.B., National Cooperative Highway Research Program, National Research Council (U.S.) (Eds.), 2004. Safety of U-turns at unsignalized median openings, NCHRP report. Transportation Research Board, Washington, D.C.

Reid, J., Sutherland, L., Ray, B., Daleiden, A., Jenior, P., Knudsen, J., Kittelson & Associates, 2014. Median u-turn intersection : informational guide. (No. FHWA-SA-14-069).

Schroeder, B., Rodegerdts, L., Bugg, Z., Jenior, P., Warchol, S., Alston, M., Haire, A., Barlow, J., Chlewicki, G., 2021. Guide for Pedestrian and Bicyclist Safety at Alternative and Other Intersections and Interchanges. NCHRP Report 948.

Schroeder, B., Warchol, S., Semensky, S., Osman, O., Ray, B., 2024a. Synthesis of Alternative Intersection Forms (No. FHWA-HRT-24-090).

Schroeder, B., Warchol, S., Semensky, S., Osman, O.A., Ray, B., Leidos, Inc., 2024b. Synthesis of Alternative Intersection Forms (No. FHWA-HRT-24-090). https://doi.org/10.21949/1521762

Shafie, N., Rahman, R., 2016. An Overview of Vehicles Lane Changing Model Development in Approaching at U-Turn Facility Road Segment. Jurnal Teknologi 78. https://doi.org/10.11113/jt.v78.9482

Shahdah, U.E., Azam, A., 2021. Safety and mobility effects of installing speed-humps within unconventional median U-turn intersections. Ain Shams Engineering Journal 12, 1451–1462. https://doi.org/10.1016/j.asej.2020.08.033

Shi, M., Tian, X., Li, X., Pan, B., 2023. The Impact of Parallel U-Turns on Urban Intersection: Evidence from Chinese Cities. Sustainability 15. https://doi.org/10.3390/su151914356

Siregar, M., Agah, H.R., Hidayatullah, F., 2018. Near-miss accident analysis for traffic safety improvement at a 'channelized' junction with U-turn. Int. J. SAFE 8, 31–38. https://doi.org/10.2495/SAFE-V8-N1-31-38

Sivaprakash, S., John, P.M., 2024. Iot Based U-Turn Vehicle Accident Prevention System (Blindends). International Advanced Research Journal in Science, Engineering and Technology 11. https://doi.org/10.17148/IARJSET.2024.11525

Stamatiadis, N., Kala, T., Clayton, A., Agent, K., 2004. U-Turns at Signalized Intersections.

Sun, X., Rahman, M.A., others, 2019a. Investigating Safety Impact of Center Line Rumble Strips, Lane Conversion, Roundabout, and J-turn Features on Louisiana Highways. Louisiana Transportation Research Center.

Sun, X., Rahman, M.A., Sun, M., 2019b. Safety Analysis of RCUT Intersection, in: 2019 6th International Conference on Models and Technologies for Intelligent Transportation Systems (MT-ITS). Presented at the 2019 6th International Conference on Models and Technologies for Intelligent Transportation Systems (MT-ITS), IEEE, Cracow, Poland, pp. 1–6. https://doi.org/10.1109/MTITS.2019.8883332

UDOT, 2011. "Thru With Delay" (web page).

Ulak, M.B., Ozguven, E.E., Karabag, H.H., Ghorbanzadeh, M., Moses, R., Dulebenets, M., 2020. Development of Safety Performance Functions for Restricted Crossing U-Turn Intersections. J. Transp. Eng., Part A: Systems 146, 04020038. https://doi.org/10.1061/JTEPBS.0000346

Wolelaw, N.M., Tessema, A.T., Alene, G.A., 2022. Modeling Behavior of U-Turning Vehicles at the Median Opening Using a Merging Behavior Approach: A Case Study in Bahir Dar City, Ethiopia. Computational intelligence and neuroscience 2022, 8273616–8273616. https://doi.org/10.1155/2022/8273616

Zubair, S., Shaikh, M.A., 2015. U-Turns and road safety — perspective from Karachi. Journal of the Pakistan Medical Association.